\documentclass[manuscript,screen]{acmart}

\usepackage[utf8]{inputenc} 
\usepackage[T1]{fontenc}    
\usepackage{url}            
\usepackage{booktabs}       
\usepackage{hyperref}
\usepackage{amsfonts}       
\usepackage{nicefrac}       
\usepackage{microtype}      
\usepackage{lipsum}
\usepackage{graphicx}
\usepackage{subfiles}
\usepackage{amsmath}
\usepackage[english]{babel}
\usepackage{amsthm}

\usepackage{listings}
\usepackage{xcolor}
\usepackage[export]{adjustbox}
\definecolor{mygray}{rgb}{0.5,0.5,0.5}
\usepackage{balance}

\setcopyright{acmcopyright}
\copyrightyear{2022}
\acmYear{2022}

\begin{document}

\title{Machine Learning Optimization of Quantum Circuit Layouts}

\author{Alexandru Paler}
\affiliation{%
  \institution{Aalto University}
  \country{Finland}}
\affiliation{%
  \institution{University of Texas at Dallas}
  \country{USA}}
\affiliation{%
  \institution{Transilvania University of Brașov}
  \country{Romania}}

\author{Lucian M. Sasu}
\affiliation{%
  \institution{Transilvania University of Brașov}
  \country{Romania}}

\author{Adrian-C\u at\u alin Florea}
\affiliation{%
  \institution{Transilvania University of Brașov}
  \country{Romania}}

\author{R\u azvan Andonie}
\affiliation{%
  \institution{Central Washington University}
  \country{USA}}
\affiliation{%
  \institution{Transilvania University of Brașov}
  \country{Romania}}

\renewcommand{\shortauthors}{Paler et al.}

\begin{abstract}
The quantum circuit layout (QCL) problem is to map a quantum circuit such that the constraints of the device are satisfied. We introduce a quantum circuit mapping heuristic, QXX, and its machine learning version, QXX-MLP. The latter infers automatically the optimal QXX parameter values such that the layed out circuit has a reduced depth. In order to speed up circuit compilation, before laying the circuits out, we are using a Gaussian function to estimate the depth of the compiled circuits. This Gaussian also informs the compiler about the circuit region that influences most the resulting circuit's depth. We present empiric evidence for the feasibility of learning the layout method using approximation. QXX and QXX-MLP open the path to feasible large scale QCL methods.
\end{abstract}


\maketitle

\section{Introduction}

The quantum circuit layout problem is deeply related to the topology of the device used to execute the circuit: instructions cannot be applied between arbitrary hardware registers. Before executing a quantum circuit, this is adapted to the device's register connectivity during a procedure called \emph{compilation}. Quantum circuit compilation is often called quantum circuit layout (QCL). The interest in efficient QCL methods is motivated by the current generation of quantum devices, called NISQ devices \cite{preskill2018quantum}. Very recent work presents worrisome evidence that even very small and shallow circuits are difficult to execute on NISQ \cite{wilson2020just}.

NISQ circuit compilation includes, for example, error-mitigation strategies \cite{endo2018practical}, flag-qubits \cite{chao2018quantum} and not just QCL methods, but the latter play definitely an significant role in the compilation of large scale fully error-corrected circuits. However, scalable compilation is not possible with current state of the art QCL methods. Fast and scalable QCL allows laying out a circuit with multiple QCL parameter values and selecting the best compiled circuit. Without going into details, our preliminary analysis showed that at the time of writing this manuscript, when optimising aggressively, compilation can take up to 1 hour for most QCL methods when presented circuits of approximately 50 qubits. It is imperative to have efficient and configurable QCL methods. We focus on three research questions:

I. How can we determine the best QCL parameter values to minimize the depth of the compiled output circuit?

II. Choosing good parameter values for the QCL method should be very fast. How can we establish a time-performance trade-off between searching optimal parameter values and the minimization of the depth?

III. Can QCL be sped up by machine learning?

More formally, QCL takes as input a circuit $C_{in}$ incompatible with a device's (can be error-corrected) register connectivity and outputs a circuit $C_{out}$ that is compatible. \emph{Additional gates} are used to overcome the connectivity limitations and compile a device-compatible circuit $C_{out}$. We define the $Ratio$ function between the depths of two circuits, where $|C|>0$ is the depth of circuit $C$. We formalize the QCL problem as follows: \emph{QCL optimization is the minimization of the $Ratio$ function.} 
\begin{align}
    Ratio(C_{in}, C_{out}) = \frac{|C_{out}|}{|C_{in}|}
\end{align}

\begin{figure}[t!]
    \centering
    \includegraphics[width=0.3\textwidth]{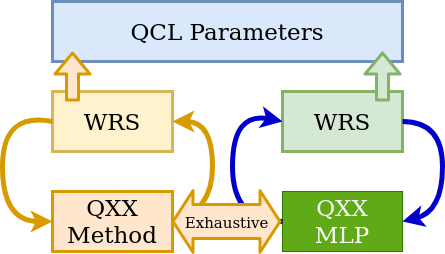}
    \caption{Continuous learning QCL: It is possible to learn our QCL method, which is called QXX (yellow), and replace it with a machine learning model (e.g. neural network - green). Optimal parameter values (blue) for adapting QXX performance are chosen by automatically executing weighted random search (WRS) in a loop.}
    \label{fig:arch}
\end{figure}

\subsection{Related Work}

QCL is already a wide topic, and we will not be providing a thorough exposition of the field. We refer the reader to the works of \cite{siraichi2018qubit, siraichi2019qubit, tan2020optimal, tan2020optimality, childs2019circuit} for detailed and careful overviews of some the works that influenced and shaped QCL.

Some of the first discussions about circuit optimality and gate counts appeared in the seminal paper of \cite{barenco1995elementary}. After quantum circuits started being analysed as reversible circuits formed from Toffoli gates, a large number of exact methods and heuristics was proposed -- a not so recent but complete review is \cite{saeedi2013synthesis}. The complexity class of QCL was discussed first in \cite{maslov2008quantum} and this has been used as a foundation for proving the complexity of different QCL variations such as \cite{siraichi2018qubit, tan2020optimal, botea2018complexity}. 

In general, there are heuristic QCL methods and exact QCL methods. A recent exact method is the one from \cite{tan2020optimal} (the paper includes a discussion about bottlenecks of exact methods). Automatic quantum circuit compilation does not always generate the best possible circuit. One of the first attempts to design full algorithm circuits considering hardware connectivity limitations is \cite{fowler2004implementation}. Another recent example of a hand optimised circuit is \cite{pallister2020jordan}, where, interestingly, the optimisation was achieved by using a results known from the automatic optimisation of circuits.

Most quantum circuit design automation tools treat QCL as a sequence of steps like initial placement and gate scheduling. The authors of \cite{siraichi2019qubit} have a complete discussion of the theoretical implications of placement and scheduling. In practice, QCL has been solved by introducing SWAP gates, but there exist more refined methods like \cite{nash2020quantum}. Solving QCL through search algorithms has been recently presented in \cite{nishio2019extracting} (beam search) and \cite{zulehner2018efficient} (A*). The work of \cite{siraichi2019qubit} introduced a parameterisable search algorithm for QCL, Bounded Mapping Tree.

In this work, we focus on QCL as an instance of register allocation as introduced by \cite{siraichi2018qubit} and use the by now classic graph view of the quantum device layout. Recently, graph-based QCL approaches (motivated by \cite{maslov2008quantum}) seem to achieve the necessary performance for compiling very large circuits to complicated device topologies. Some QCL approaches were proposed using hyper-graphs \cite{andres2019automated}, but more classical approaches are the ones from \cite{duncan2020graph}.

Machine learning has started being widely applied in different aspects of quantum computing. Nevertheless, compared to \cite{zhang2020topological} we are not compiling only single qubit gates after needing impractical numbers of hours to train a very large network. In parallel and independent to the preparation of this manuscript, machine learning methods for QCL have been presented in \cite{pozzi2020using}. Compared to \cite{pozzi2020using}, our method is capable of learning continuously.

QCL includes two steps \cite{siraichi2019qubit}. The first step is \emph{initial placement} (e.g., \cite{paler2019influence}), where a mapping of the circuit qubits to device registers is computed. This step is also called qubit allocation \cite{siraichi2018qubit}. The second QCL step is \emph{gate scheduling}, where the circuit $C_{in}$ is traversed gate-by-gate.

\begin{figure}[t!]
\centering
\includegraphics[width=6cm]{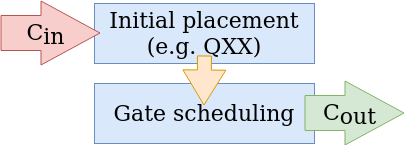}
\caption{The quantum circuit compilation (QCL) procedure takes an input circuit $C_{in}$ and transforms it into the functionally equivalent $C_{out}$ circuit. The QXX method computes an optimal assignment of circuit qubits (registers) to device qubits (registers).}
\label{fig:procedure}
\end{figure}

\subsection{Contribution}

We present machine learning method (cf. Fig.~\ref{fig:arch}) for the initial placement (i.e. mapping) of quantum circuits. In order to test the feasibility of learning, we develop an initial placement heuristic, which we called QXX for no particular reason.

From a technical perspective, our methods are designed and implemented into great detail and capable of being deployed in practice. The machine learning model of QXX is called QXX-MLP, which is the significantly faster implementation of the QXX. At the date of writing this manuscript, our method was the first to effectively learn a QCL heuristic.

QXX is configurable over multiple parameters, and our goal is to have a method that can automatically chose the best parameter values in order to achieve optimally compiled circuits. The novelties of QXX and QXX-MLP (Section~\ref{sec:methods}) are:
\begin{enumerate}
    \item automatic feature selection -- focusing on the important sub-circuit using a configurable Gaussian function (Section~\ref{sec:gdepth});
    \item automatic QCL configuration -- we use weighted random search (Section~\ref{sec:wrs}) to optimize QXX parameter values. The parameters influence also the speed of the QCL;
    \item demonstrating QCL learnability -- the machine learning version QXX-MLP works as an approximation method for the circuit layout depth (Section~\ref{sec:learning});
    \item scalability -- the almost instantaneous layout performance (Sections~\ref{sec:discmlp} and ~\ref{sec:timeouts} and Table~\ref{tbl:wrsbest}). We show that QCL execution time can be shortened while maintaining the $Ratio$ optimality.
\end{enumerate}

The goal of this work is to highlight the generality and wide applicability of learning QCL methods. For this reason, we focus on benchmarking circuits which are compatible with both non-error-corrected and error-corrected machines. We use synthetic benchmarking circuits, QUEKO\cite{tan2020optimality} (Section~\ref{sec:bench}), which capture the properties of Toffoli+H (e.g. arithmetic, quantum chemistry, quantum finance etc.) circuits without being specific for a particular application. Section~\ref{sec:results} presents and discusses experimental results performed with the QXX and QXX-MLP approaches. Conclusions are synthesized in Section \ref{sec:conclusions}.

\section{Methods}
\label{sec:methods}

We present the QXX method and describe the machine learning techniques that are using it. Subsection \ref{sec:initial} gives the details of QXX. Fig.~\ref{fig:arch} illustrates the approach followed by this work during the QCL parameter optimization stage: The parameters of the normal QXX method (orange) are optimized using WRS. To further speed up QXX we train a machine learning model, called QXX-MLP, to predict the $Ratio$ values obtained by using the QXX for a given circuit and a particular set of parameter values.

We employ three methods to evaluate how the parameter optimization of QXX influences the compiled circuits. First, an efficient parameter optimizer is Weighted Random Search (WRS, Subsection \ref{sec:wrs}). Second, we use exhaustive search to collect training data to obtain the QXX-MLP neural network (Subsection \ref{sec:learning}). The latter is used to estimate optimal parameter values.

The WRS method starts from an initial set of parameter values and adapts the values in order to minimize the obtained $Ratio$. This procedure forms a feedback loop between WRS and the QXX method. Running WRS multiple times for a set of benchmark circuits is equivalent to \emph{almost} executing an exhaustive search of the parameter space. The exhaustive search data is collected and used to train QXX-MLP.

From a methodological point of view, device variabilities are very important and, moreover, these seem to have a fluctuating behavior \cite{wilson2020just}. However, in this work we do not consider that NISQ qubits have variable fidelities \cite{tannu2019not}, or that crosstalk is a concern during NISQ compilation \cite{murali2020software}.

We will use the following notation. The first QCL step computes a list $M$, where $M[q]=r$ refers to circuit qubit $q$ being stored on device register $r$. Computing the list $M$ is the analogue to determining a good starting point for the gate scheduling procedure. We represent two-qubit gates by the tuples $(q_i, q_j)$. Scheduling executes the current two-qubit gate if $(M[q_i], M[q_j])$ is an edge of device connectivity graph, which will be called \emph{DEVICE}. Otherwise, the gate qubits are moved across the device and stored in registers connected by an an edge from \emph{DEVICE}. The movement introduces additional gates, such as SWAP gates, in order for all tuples $(q^{out}_i, q^{out}_j)$ to be edges of \emph{DEVICE}. The mapping is updated accordingly, as illustrated in Fig.~\ref{fig:move}. In general, the depth of $C_{out}$ is lower bounded by $|C_{in}|$.

\subsection{The QXX Mapping Heuristic}
\label{sec:initial}

QXX is a fast search algorithm to determine a qubit mapping (allocation), and is called by the subsequent gate scheduler to compute a good qubit allocation/mapping/placement. QXX  uses an estimation function to predict how a $C_{out}$ with minimum depth would have to be initially mapped.

A novelty is that QXX uses a Gaussian-like function called $GDepth$ to estimate the resulting depth (cost) of the layed out circuit called  $C_{out}$. The qubit mapping is found using the minimum estimated value of $GDepth$. QXX uses \emph{three types of parameters}, which we will explain in the following three sections:
\begin{enumerate}
    \item for configuring the search space;
    \item for adapting $GDepth$ to the circuit $C_{in}$;
    \item for adapting QXX to the second step of QCL, namely the scheduler/router of gates.
\end{enumerate}

\subsection{Search Space Configuration}
\label{sec:space}

\begin{figure}[t!]
\centering
\includegraphics[width=7cm]{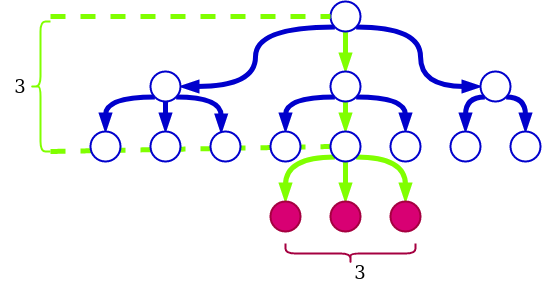}
\caption{The search space dimensions of QXX can be configured by adapting the parameters \emph{MaxDepth} (green) and \emph{MaxChildren} (red). Each node of the search tree stores a list of possible mappings for which a minimum cost was computed. The maximum length of the list is \emph{MaxChildren} (e.g., 3 blue levels). The list is emptied if a lower minimum cost is computed, or if \emph{MaxDepth} has been achieved. In the latter case, the path with the minimum cost (green) is kept, and all other nodes are removed.}
\label{fig:tree}
\end{figure}

QXX is a combination of breadth-first search and beam search. The search space is a tree (cf. Fig.~\ref{fig:tree}). Constructing a qubit-to-register mapping is an iterative approach: qubits are selected one after the other, and so are the registers where the qubits mapped initially. 

For example, consider $C_{in}=\{(q_1,q_2), (q_2,q_3), (q_3,q_4)\}$ a circuit of three CNOTs and a mapping $M=[r_1,r_2]$. This means that $q_1$ is allocated to register $r_1$, and $q_2$ to $r_2$. After the first qubit was mapped, we have $|M|=1$. After all qubits were mapped, we have $|M|=Q$. The maximum depth of the tree is $Q$. In the worst case, each node has $Q$ children.

The tree is augmented one step at a time, by adding a new circuit qubit $a$ to the mapping (in the order of their index, in the current version - we do not analyze the influence of this choice). This increases the tree's depth: at each existing leaf node, all possible $N$ mappings of $a$ are considered. Consequently, all the new leaves of a tree are the result of appending $a$ to the previous' level leaves, which now are usual nodes.

New depth estimation values are computed using the $GDepth$ function, each time leaves are added to the tree. Each tree node has an associated $GDepth$ cost. The level in the tree equals the length of the mapping for which the cost was computed.

The search is stopped after computing a complete mapping with the minimum $GDepth$ cost. Thus, the maximum number of leaves per node is added in the unlikely case that all values of $GDepth$ are equal.

The search space will easily explode for large circuits. We introduce two parameters to prune the search space. In Fig.~\ref{fig:tree}, the result of pruning the search space tree is represented by the green path and the green bounding box.

The first parameter is $MaxChildren$ whose job is to limit the number of children of equal minimum $GDepth$ values. For an arbitrary value of $MaxChildren < Q$, the tree will include at level $l$ at most $l \cdot MaxChildren$ nodes. 

The second parameter is the cut-off threshold $MaxDepth$ which specifies that, at levels indexed by multiples of $MaxDepth$, all the nodes are removed from the tree, except for the ancestors of the minimum cost leaf. This is because for large circuits (e.g. more than 50 qubits) it is not practical to evaluate all the new combinations of $Q$ registers for $l\cdot MaxChildren$ leaves.

\subsection{Configuration of Circuit Depth Estimation}
\label{sec:gdepth}

The $GDepth$ function is used to estimate the depth of the compiled circuit without laying it out. 

The value of $GDepth$ can be calculated once at least two qubits were mapped. Equivalently, the value of $GDepth$ is computed only for CNOTs whose qubits were mapped. The $GDepth$ function is a sum of a Gaussian functions whose goal is to model the importance of CNOT sub-circuits from $C_{in}$. In other words, $GDepth$ is the sum of the costs estimated for scheduling the CNOTs of a circuit:

{\small
\begin{align}
    GDepth  &= \sum_{i=0}^{N_c} dist_{i} \exp{\left(-B \cdot\left\Vert \frac{i}{N_c} - C\right\Vert^2 \right)}\\
    &\text{ where } N_c \leq |C| \nonumber
\end{align}
}

In the above formula, $N_c$ is the number of CNOTs from $C_{in}$ whose qubits were already mapped. The $B$ and $C$ parameters control the spread and position of the Gaussian function. The value of $i$ is the index of the CNOT from the resulting circuit and $dist_i$ is the cost of moving the qubits of the CNOT. 

For example, for the circuit $$C_{in}=\{(q_1,q_2), (q_2,q_3), (q_3,q_4)\}$$ and the mapping $M=[r_1, r_2]$, $N_c=1$ the $GDepth$ can be computed for only the first CNOT, because $q_3$ and $q_4$ were not mapped.  .

The scheduled CNOTs are indexed, and \emph{we assume that all gates are sequential} (parallel gates are executed sequentially, considering the circuit truly a gate list). For example, after scheduling two CNOTs of an hypothetical circuit, the two CNOTs are numbered $1$ and $2$. In the exponent of $GDepth$ the value of $\frac{i}{N_c}$ is always in the range $[0,1]$. 

For $dist_i$ we use the shortest distance on an undirected graph. An example is Fig.~\ref{fig:move}, where $dist_i = 5$. For two qubits $a,b$, where  $(M[a], M[b])$ is already an edge of \emph{DEVICE}, the $dist_i = 1$. Otherwise, the distance between two qubits is computed based on the edge weights attached to the \emph{DEVICE} graph.

\begin{figure}[h!]
\centering
\includegraphics[width=12cm]{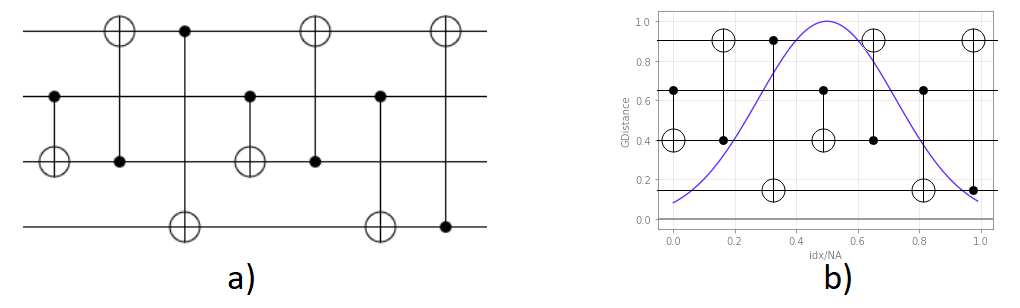}
\caption{A quantum circuit and a Gaussian function. a) A four qubit quantum circuit consisting of a sequence of CNOTs; b) The $dist_i$ uses a Gaussian function. The latter is drawn superimposed in order to highlight the weight generated by the Gaussian at each CNOT position in the circuit. The weights are minimal for the first and last CNOT. The maximum value is for the CNOT at the middle of the circuit.}
\label{fig:gaussweight}
\end{figure}

\begin{figure}[h!]
\centering
\includegraphics[width=8cm]{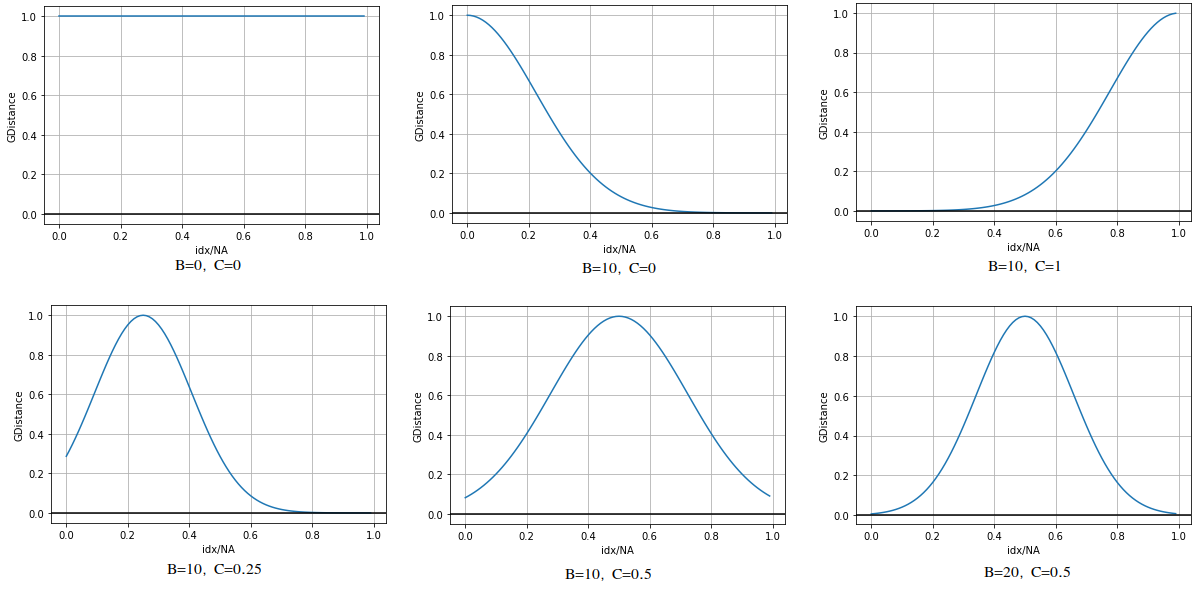}
\caption{The value of the Gaussian used by $GDepth$ for different values of parameters $B$ and $C$. The horizontal axis illustrates the input value to the Gaussian: the gates' integer index in the circuit is scaled with the total number of gates in the circuit. The importance (weight) of a gate is highest whenever the value of the Gaussian approaches the maximum value. A gate has low importance when the Gaussian has low values.}
\label{fig:gauss}
\end{figure}

If the intention is to allow CNOTs at the middle of the circuit to have longer movements on the device, we can set parameters to generate a function like in the top right panel of Fig.~\ref{fig:gauss}. For example, for $B=5$ and $C=0.5$, the $dist_i$ from $GDepth$ are weighted with almost zero at the start of the circuit (Fig.~\ref{fig:gaussweight}). The opposite situation is illustrated in the middle panel. For $B=0$, the Gaussian is effectively a constant function, such that the cost is the sum of all the CNOT distances. 

\subsection{Configuring QXX for the Scheduler}
\label{sec:adapting}

QXX is effectively estimating the output of the gate scheduler, which \emph{selects the best edge} of \emph{DEVICE} where to execute the CNOTs of a circuit. The scheduler is modeled as a black box and its functionality is unknown. QXX can work with Qiskit's StochasticSwap, tket or other tools such as the ones ones from \cite{li2019tackling, zhang2021time}.

Starting from the initial placement, QXX estimates the total depth of the circuit after repeatedly mapping the circuit. Without loss of generality, QXX assumes that the scheduler will move both qubits of a CNOT across the \emph{DEVICE} towards the selected edge. Instead of updating the mapping, the movements of the qubits are \emph{accumulated} into an \emph{offset} variable.

Qubit movements is captured by the $MovementFactor$ parameter. The $MovementFactor$ is asymmetric, and when processing CNOT qubits, it moves on the \emph{DEVICE} the qubit with the lowest index by the fraction $\frac{1}{MovementFactor}$, and the qubit with the highest index by $\frac{MovementFactor - 1}{MovementFactor}$.

As we will show in the Results section, for deep circuits we determine $MovementFactor$ values closer to 2 (favors high and low index qubits equally), and for shallow circuits the values are higher (favors low indexed qubits).

\begin{figure}[h!]
\centering
\includegraphics[width=4cm]{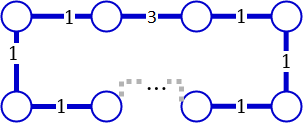}
\caption{Device connectivity graph is blue, the weight of all but one edge is 1. For example, the high cost might be due to higher physical error-rates.}
\label{fig:cost}
\end{figure}

After moving the qubits on the device (Fig.~\ref{fig:cost}), the offset of a qubit is an estimation of how much the qubit was moved by the scheduler. For example, the offset of an arbitrary qubit used in $3$ CNOTs is the sum of the three movement updates which are obtained after scaling each CNOT's $dist_i$ with the corresponding $MovementFactor$ expression.

\begin{figure}[h!]
\centering
\includegraphics[width=5cm]{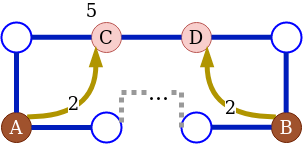}
\caption{The effect of the \emph{MovementFactor}: Device connectivity graph is blue, and the two brown qubits A and B have to be interacted. If not specified, all edges have weight 1. The shortest path between A and B has cost L=5, because there are five edges between A and B. There are multiple options how A and B can be brought together. One of the options is to assume that A and B move to the pink C and D. Assuming that CD is at the middle of the path connecting A and B, then \emph{MovementFactor=2} because both A and B are moved on average L/2. A higher movement factor implies that one of the qubits moves less, while the other more.}
\label{fig:move}
\end{figure}

The movement of qubits on the device is controlled by an additional parameter, called $EdgeCost$: higher values imply a larger estimation of the movement heuristic. Changing the value of $EdgeCost$ is equivalent to scaling the total value of $GDepth$, because $EdgeCost$ is a common factor in the calculation of $dist_i$. For the weight scale factor $EdgeCost=1$, as shown in Fig.~\ref{fig:move}, the minimum distance between the CNOT qubits corresponds to the sum of the edge weights separating them (five in Fig.~\ref{fig:move}).

\begin{table}[t]
\small
\centering
\begin{tabular}{ l r r r | r r | r r r}
     & \multicolumn{3}{c}{QXX Parameters} & \multicolumn{2}{c}{WRS Obtained}  & \multicolumn{3}{c}{Exhaustive Search}\\
Name & Min & Max & Increment & WRS-Weight& Prob. of Change & Start   & Stop   & Increment\\
\hline
MaxDepth        & 1 & 55    & 1     &   9.35    &   0.62    & 1     & 9     & 4\\
MaxChildren     & 1 & 55    & 1     &   8.00    &   0.53    & 1     & 9     & 4 \\
B               & 0 & 500   & 0.1   &   7.76    &   0.52    & 0     & 20    & 2 \\
C               & 0 & 1     & 0.01  &   15.06   &   1.00    & 0     & 1     & 0.25\\
MovementFactor  & 1 & 55    & 1     &   3.52    &   0.23    & 2     & 10    & 4 \\
EdgeCost        & 0.1 & 1   & 0.1   &  10.59    &   0.70    & 0.2   & 1     & 0.4
\end{tabular}
\caption{Parameter values. The \emph{QXX Parameters} columns represent the min. and max. value together with the increments be to used with the heuristic. The \emph{WRS Obtained} columns illustrate the results of the weighted random search method for optimal parameter values. Ranges and increments of QXX's input parameters. The training of QXX-MLP as well as WRS were performed for a smaller search space whose parameter ranges are presented in the \emph{Exhaustive Search} columns.}
\label{tbl:params}
\end{table}

\subsection{Weighted Random Search for Parameter Configuration}
\label{sec:wrs}

The QXX parameters from Sections~\ref{sec:space},~\ref{sec:gdepth},~\ref{sec:adapting} have to be tuned for optimal performance of the compilation. We call \textit{trial} the evaluation for one set of parameter values. In previous work \cite{Florea2019}, we introduced the WRS method \footnote{https://github.com/acflorea/goptim}, a combination of Random Search (RS) and probabilistic greedy heuristic. Instead of a blind RS search, the WRS method uses information from previous trials to guide the search process toward next interesting trials. We use WRS to optimize the following QXX parameters (introduced in Subsection \ref{sec:initial}): $MaxDepth$, $MaxChildren$, $B$, $C$, $MovementFactor$, and $EdgeCost$ (see Table \ref{tbl:params}).

Within the same number of trials, different optimization methods achieve different scores, depending on how ``smart'' they are. Due to the nature of QXX, the trial execution times are variable, since it depends on the defined quantum architecture, the topology of the circuit to lay out, as well as the values of the parameters (cf. Fig.~\ref{fig:tree}). Search space reduction and search strategy are inter-connected. For the exhaustive search parameter ranges from Table \ref{tbl:params}, we limit the time spent evaluating a parameter configuration by introducing a \emph{timeout} parameter. WRS uses the obtained data to run an instance of fANOVA \cite{HutHooLey14}.

\subsection{Learning QXX - Training QXX-MLP}
\label{sec:learning}

The previous section was about searching parameter values. Herein, we go one step further and learn the behavior of the gate scheduler in relation to the mapper parameters and the $GDepth$ function used by QXX. We describe the method to learn specific tuples consisting of circuit and QXX parameters, and how to estimate the depth of the mapped circuits.

Parameter optimization, and for that reason, efficient initial mapping of the circuit to the device is a regression problem. The training data for learning was obtained as follows: for the parameters from Table~\ref{tbl:params}, we choose smaller ranges and larger increments as illustrated in the \emph{Exhaustive Search} columns. There are three possible values for $MaxDepth$ and three values for $MaxChildren$. For a given value of $MaxDepth$, there are $3 \times 11 \times 5 \times 3 \times 3 = 1485$ possible parameter configurations. There is a total of $90 \times 1485 = 133650$ layouts for each  $MaxDepth$. Overall, there are $400950$ parameter configurations of the 90 circuits that are being evaluated.

Table \ref{tbl:params} illustrates the parameter ranges for which we collected data that allows us to compute the ratio $Ratio(C_{in}, C_{out})$ for every combination of circuit layout and QXX parameters. The generated exhaustive search data is available in the project's online repository.

We considered three candidate models to learn QXX: k Nearest Neighbors (KNN), Random Forest (RF) and MultiLayer Perceptron (MLP). Each model has different inductive bias \cite{Mitchell80}, being respectively: a local--based predictor, an ensemble model built by bootstrapping, and a connectionist model, respectively. 

Despite of their conceptual simplicity, KNN predictors are easily interpretable and passed the test of time \cite{Wu2007}. RFs were found as the best models for classification problems \cite{Delgado2014}, and we wanted to investigate their performance on this regression problem as well. Finally, MLPs creates new features through nonlinear input feature transformations, unlike KNN and RF which use raw input attributes. Nonlinear transformations and non-local character of MLPs are considered the premises for the successful deep learning movement \cite{Bengio2009}.

We found MLP as the best model and use it during the parameter optimization stage (as illustrated in Fig.~\ref{fig:arch}) as an approximator for the functionality of QXX. More details are in the Appendix.

\section{Results}
\label {sec:results}

We present empirically obtained results about: 1) the performance of QXX; 2) the quality of the QXX-MLP model; 3) the performance of optimising QXX parameters with WRS. We select QXX parameter values using: a) exhaustive search; b) WRS with the QXX method (orange in Fig.~\ref{fig:arch}) , and c) WRS on the QXX-MLP (green in Fig.~\ref{fig:arch}).

The QXX method was implemented and is available online\footnote{\url{https://github.com/alexandrupaler/qxx}}. The current implementation is agnostic of the underlying quantum circuit design framework (e.g. Cirq or Qiskit). QXX is implemented in Python. The exhaustive search was executed on an i7 7700K machine with 32GB of RAM. The QXX model was trained on Intel Xeon W-2145 3.70GHz with 16 cores and 256 GB RAM. The WRS parameter optimization was performed on a laptop grade i5 processor with 16 GB RAM. For the benchmarks and comparisons we used Cirq 0.9, IBM Qiskit 0.25 and tket 0.2.

In the following, we describe how the WRS heuristic is used to evaluate the QXX method and its MLP implementation. Afterwards, we present a series of plots that support empirically the performance of QXX. We analyze the influence of the parameters, and offer strong evidence in favor of learning quantum circuit layout methods. In particular, we will show that the $GDepth$ Gaussian can shorten the time of necessary to run QXX. This is achieved by pruning the search space and focusing on the most important region of the circuit. We will present examples for how the Gaussian is automatically adapted for deep and shallow circuits.

\subsection{Benchmark Circuits}
\label{sec:bench}

We evaluate the $Ratio$ fitness of the QXX method using the QUEKO benchmark suite \cite{tan2020optimality}. These circuits abstract Toffoli based and quantum supremacy like circuits, as well as a variety of NISQ chip layouts. Such benchmarks complement the libraries of reversible adders and quantum algorithms \cite{li2020qasmbench}.

Automatic compilation/mapping of circuits, although applicable to NISQ applications, e.g. Fig~\ref{fig:timeout_qse}, is of little practical importance by itself when one wishes fault-tolerance to be taken into account. NISQ circuits, such as for supremacy or for VQE/QAOA, are co-designed and a method like ours is just one piece in a much larger workflow which. We focused on Toffoli+H circuits, QUEKO TFL, because such circuits are not co-design and are also very representative for large scale error-corrected computations.

The 90 QUEKO TFL circuits include circuits with known optimal depths of $[5, 10, 15, 20, 25, 30, 35, 40, 45]$ (meaning that the input circuit and the layout circuit have equal depths $|C_{out}|=|C_{in}|$). For each depth value there are 10 circuits with 16 qubits. The NISQ machine to map the circuits to is Rigetti Aspen. A perfect QCL method will achieve $Ratio=1$ on the QUEKO benchmarks.

In our experimental setup, the layout procedure uses: 1)
QXX for the initial placement (first QCL step) and 2) the Qiskit \texttt{StochasticSwap} gate scheduling (second QCL step). The results depend on both the initial placement as well as the performance of the \texttt{StochasticSwap} scheduler. We do not configure the latter and use the same randomization seed. We assume that this is the reason why the QXX performance is close to the Qiskit one.

\begin{figure}[t!]
\centering
\includegraphics[width=8cm]{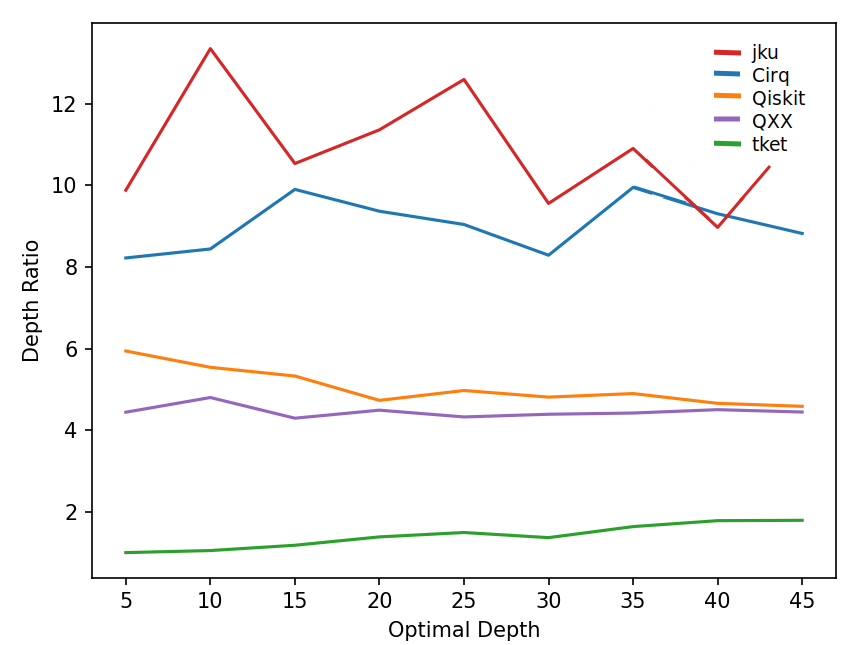}
\caption{Comparison with state of the art methods using QUEKO TFL benchmark circuits.  All curves except QXX are from \cite{tan2020optimality}. The curves are computed after averaging the depth ratios for 10 circuits for each known optimal depth from the benchmark. Horizontal axis is the known optimal depth of the TFL circuits. The vertical axis illustrates the achieved depth $Ratio$. We have used QXX together with the Qiskit scheduler, and we assume that this is the reason why the QXX curve is close to the Qiskit curve. For shallow circuits the mapper is more important than the scheduler (QXX is better than Qiskit), and for deeper circuits the importance of the mapper vanishes (QXX and Qiskit perform close to each other). QXX is not a scheduler, but a mapper (cf. Section~\ref{sec:adapting}).}
\label{fig:total}
\end{figure}

\subsection{Resulting Depths and Scalability}
\label{sec:timeouts}

Our goal is to show that the behavior of the mapper/compiler can be learned, and that compilation can be sped up using machine learning.  QXX achieves $Ratio$ values around 30\% lower (which means better -- best $Ratio$ is 1) than Qiskit on the low depth (up to 15 gates) QUEKO TFL circuits. In general, as shown in Fig.~\ref{fig:total}, the performance of QXX is between Qiskit and tket \cite{sivarajah2020t}. tket outperforms Qiskit, and QXX too, on the QUEKO benchmarks, because it has a much smarter scheduler. QXX is not a scheduler, but a mapper (cf. Section~\ref{sec:adapting}).

The results from Fig.~\ref{fig:total} are encouraging, because: a) QXX performs better than most compilers; b) there is known variability in the compiler's performance with respect to benchmark circuits, such that for other circuits the classification might look completely different; c) our results were obtained very fast when using the MLP approach -- we get almost instantaneously the layed out circuit (Section~\ref{sec:discmlp}).

For NISQ gate error rates (realistically) upper bounded by $10^{-2}$, only circuits with a maximum depth of 30 are of practical importance.  For shallow circuits the mapper is more important than the scheduler (QXX is better than Qiskit), and for deeper circuits the importance of the mapper vanishes (QXX and Qiskit perform close to each other). Fig.~\ref{fig:timeout_tfl25} presents results with parameters chosen specifically for shallow circuits.

With respect to the scalability of our methods, one interesting aspect are the timeouts. For extreme parameter values (e.g. when $MaxChildren$ and $MaxDepth$ are $9$) the execution time of QXX is high although the method has polynomial complexity. We introduced a timeout of 20 seconds for the QXX executions and collected data accordingly. Table~\ref{tbl:timeouts} illustrates the increasing execution times, it offers the motivation to learn the method -- the model will have constant execution time irrespective of the parameter value configuration.

Additionally, we notice that computing a good initial placement takes, in the best case, a small fraction of total time spent laying out. Without considering the optimality of the generated circuits, in the worst case, computing the initial placement using QXX can take between 2\% and 99\% of the total layout time. For more details see Table~\ref{tbl:wrsbest} and Section~\ref{sec:discmlp}. For $MaxDepth=1$ the maximum time fraction is 10\%, for $MaxDepth=5$ the maximum is 85\%, while for $MaxDepth=9$ it is 99\%. These values are also in accordance with the execution times presented in Table~\ref{tbl:timeouts}. However, when considering the 100 fastest QXX execution times, for each of the $MaxDepth$ values, the maximum mapping duration is 4\% of the total layout duration.

\begin{table}[t!]
    \small
    \centering
    \setlength{\tabcolsep}{4pt} 
    \begin{tabular}{r r | r r r r}
    \multicolumn{2}{c}{Search Space Params.}& \multicolumn{4}{c}{Timeout}\\
    MaxDepth    & MaxChildren   &   0.05s    &   0.5s &   5s   &   20s\\
    \hline
    1           &   1           &   25      &   0	&   0	&   0\\
    1           &	5           &	43	    &   0	&   0	&   0\\
    1	        &   9	        &   36      &   3   &   0   &   0\\
    5	        &   1           &   25      &   0   &   0   &   0\\
    5	        &   5           &   23899   &   0   &   0   &   0\\
    5	        &   9	        &   41014   &   2457&	0	&   0\\
    9	        &   1           &   46	    &   0	&   0	&   0\\
    9	        &   5       	&   44550   & 37365 & 2501  &   172\\
    9	        &   9           &	44550   &  44494& 36288 &   28798
    \end{tabular}
    \caption{The number of WRS timeouts (0.05s, 0.5s, 5s, 20s) is a measure of QCL execution time. We count the number of timeouts when using different search space pruning strategies configured by $MaxDepth$ and $MaxChildren$.}
    \label{tbl:timeouts}
\end{table}

\subsection{QXX parameter optimization using WRS} 
\label{sec:paramopt}

Initially we ran WRS for a total of 1500 trials within the parameters space defined in Table~\ref{tbl:params}. For $MaxDepth$, $MaxChildren$ and $MovementFactor$ we use the same limits and steps defined in the table. For $B$, $C$, and $EdgeCost$ we generated the values by drawing from a uniform distribution in the specified range. 

We ran the classical RS step for 550 trials and computed the weight (importance) of each of the parameters using fANOVA obtaining the values from Table~\ref{tbl:params}. The weight of a parameter measures its importance for the optimization of the fitness function.

We use the $mean(Ratio)=mean(|C_{out}|/|C_{in}|)$ fitness measure. WRS ran with eight workers for a total time of four hours and 11 minutes and the best result (3.99) was obtained at iteration 1391 and again, at a later trial, for a different value of $MaxDepth$. Table~\ref{tbl:wrsbest} shows the combination of parameter values that yield the best results.

The subsequent WRS executions use the exhaustive search parameter space defined in Table~\ref{tbl:params}. In this parameter space we use WRS to optimize several configurations for which we have changed either the evaluation timeout with values from Table~\ref{tbl:timeouts} or the maximum TFL Depth (either 25 of 45). Interestingly, the 5 seconds timeout, achieves a better performance than the WRS parameter optimization with 20 seconds timeout (cf. large number of timeouts in Table~\ref{tbl:timeouts}).

In general, WRS selects high $MaxDepth$ values. This is explainable by the fact that optimal $Ratio$ values are easier to be found using large search spaces.

\begin{table}[t]
\small
\centering
\setlength{\tabcolsep}{2pt} 
\begin{tabular}{ l | r r r r  | r r r r | r}
            & \multicolumn{4}{c|}{TFL-45}   & \multicolumn{4}{c |}{TFL-25}  & MLP   \\
 Name       & 20s  & 5s   & 0.5s    & 0.05s     & 20s  & 5s   & 0.5s    & 0.05s     \\
\hline
MaxDepth    & 9     & 9     & 9     & 6     &  8    & 9     & 9     & 3     & 9     \\
MaxChild    & 4     & 3     & 2     & 2     &  4    & 3     & 2     & 2     & 9     \\
B           & 5     & 17.3  & 8     & 3.5   & 6.9   & 15.7  & 6.10  & 2     & 1.5   \\
C           & 0.61  & 0.25  & 0.02  & 0.31  & 0.86  & 0.91  & 0.65  & 0.74  & 0.32  \\
Mov.Factor  & 2     & 4     & 2     & 6     & 6     & 10    & 6     & 7     & 10    \\
EdgeCost    & 0.2   & 0.2   & 0.2   & 0.9   & 0.2   & 0.2   & 1     & 0.2   & 0.8 \\
\hline
Duration(s) & 35170 & 28800 & 21785 & 11438 & 10583 & 8110 & 6497 & 4102 & \textasciitilde 2\\
Avg. Ratio  & 4.093 & 4.138 & 4.328 & 4.465 & 4.043 & & 3.966 & 4.279 4.537 & 4.423
\end{tabular}
\caption{Parameter values obtained using WRS on the batch of 90 QUEKO circuits. The TFL-45 and TFL-25 are for QXX and WRS on the the TFL circuits of depths 25 and 45. The MLP column is for when using QXX-MLP with WRS. The last two rows represent the durations and the recorded average Ratios. The MLP approach is very fast and takes approx. 2 seconds.}
\label{tbl:wrsbest}
\end{table}

\subsection{QXX-MLP using WRS: Fast and scalable QCL}
\label{sec:discmlp}

\begin{figure}
\centering
\includegraphics[width=8cm]{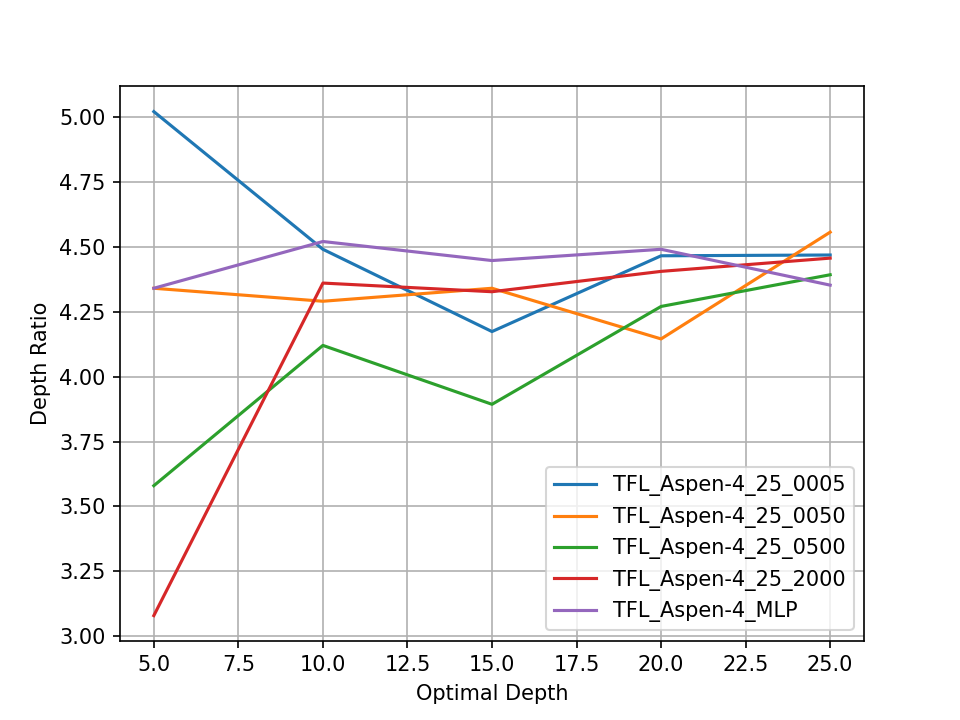}
\caption{QXX-MLP achieves approximately 90\% of the QXX performance when laying out TFL circuits with depths up to 25 -- the most compatible with current NISQ devices. WRS was applied on: 1) the normal QXX and timing out too long evaluations; 2) the QXX-MLP model.}.
\label{fig:timeout_tfl25}
\end{figure}

One of the research questions was if it is feasible to learn the QCL methods in general, and QXX in particular. Moreover, 
we answer the question ``Would it be possible to choose optimal values by a rule of thumb instead of searching for them?''. For example, is it possible to obtain good $Ratio$ in a timely manner by using low $MaxDepth$ values? During our experiments, the found parameters did not differ significantly between QXX and QXX-MLP. The results from Fig.~\ref{fig:timeout_tfl25} and Table~\ref{tbl:wrsbest} show that the MLP model of QXX performs well -- within 10\% performance decrease compared to the normal QXX.

We used a combination of WRS and QXX-MLP in an attempt to minimize the time required to identify an optimal configuration. QXX-MLP and WRS are almost instantaneous: under $2$ seconds for the entire batch of 90 circuits. In contrast, the execution time of WRS using the normal QXX was about 3 hours to find the optimal parameters for the 90 circuits. A detailed analysis of the speed/optimality tradeoffs available in the Appendix (e.g. Fig.~\ref{fig:resdepth}).

\begin{figure}[t!]
\centering
\includegraphics[width=8cm]{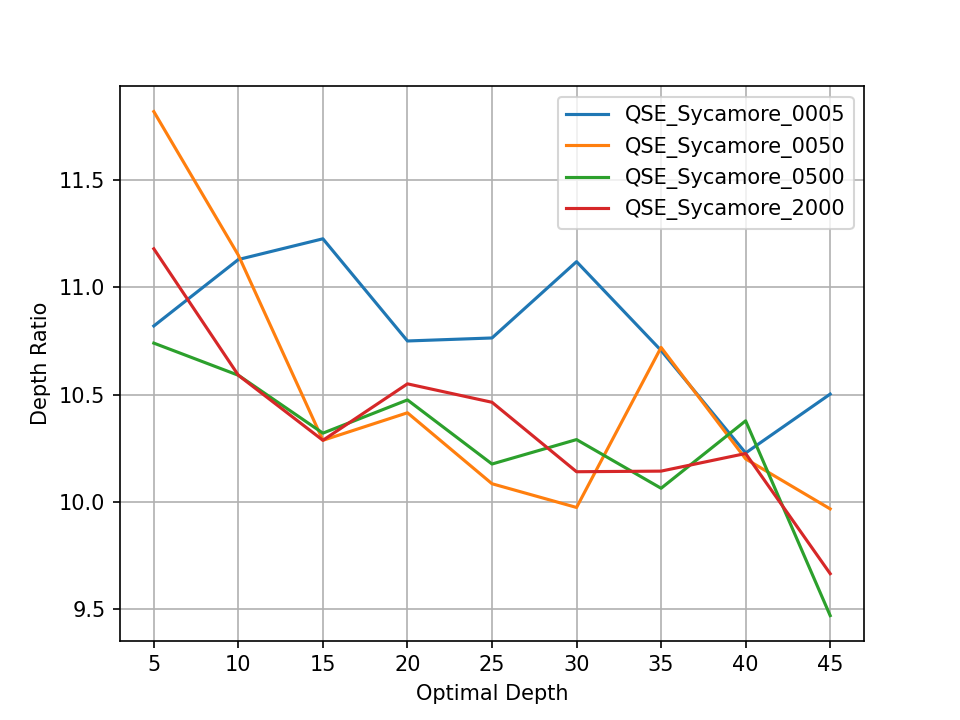}
\caption{Transfer of learning: Laying out QUEKO QSE (supremacy experiment) circuits using parameters learned from QUEKO TFL circuits. Each parameter evaluation executed by WRS was timed out after 5, 50, 500 and 2000 $1/100th$ seconds (10 milliseconds).}
\label{fig:timeout_qse}
\end{figure}

We also tested the \emph{transfer of learning}: the possibility of training QXX on a set of circuits, and then applying to a different type of circuits. Figure~\ref{fig:timeout_qse} illustrates the results of applying QXX-MLP on quantum supremacy circuits (QSE)\cite{tan2020optimal}.

The performance of QXX-MLP is more than encouraging: it had the performance of a timedout WRS optimization. The WRS parameter optimization did not timeout, because MLP inference is a very fast constant time operation. Table~\ref{tbl:wrsbest} shows that QXX-MLP performs similarly to the WRS with a 0.05s timeout. For example, when applied to QXX-MLP, the values chosen by WRS for $EdgeCost$ and $MovementFactor$ \emph{are consistent} with the exhaustive search results (cf. Fig.~\ref{fig:rescx} in Appendix): in general, an $EdgeCost=0.2$ is preferable for all the TFL-depth values, and for $MovementFactor > 2$ is preferable. The preference for large movement factors is obvious for the shallow TFL circuits. 

Regarding QCL speed: The WRS and QXX-MLP optimization takes a few seconds compared to the hours (cf. Table~\ref{tbl:wrsbest} for laying out all the circuits from Section~\ref{sec:bench}) necessary for WRS and QXX. This is a great advantage that comes on the cost of obtaining a trained MLP, which is roughly the same order of magnitude to a WRS parameter optimization. MLP training is performed only once. In the case where the MLP model is not used WRS parameter optimization has to be repeated. In a setting where quantum circuits are permanently layed out and executed (like in quantum computing clouds) incremental online learning is a feasible option. 

The parameter optimization of QXX-MLP seems to be very predictable, because the corresponding curves in Fig.~\ref{fig:timeout_tfl25} are almost flat. Table~\ref{tbl:wrsbest} provides evidence that the MLP is very conservative with the choice of the values of $B$ and $C$: The Gauss curve is almost flat, and positioned slightly towards the beginning of the circuit. The flatness of the curve and its position could explain the almost constant performance from Fig.~\ref{fig:timeout_tfl25} and the 10\% average performance degradation of QXX-MLP.

\subsection{Automatic Subcircuit Selection}

Our results support the thesis that WRS can adapt the parameters of the QXX Gauss bell curve in order to select the region of the circuit that influences the total cost of laying it out. The parameter controlling the center of the bell curve is relevant with respect to resulting layout optimality, as well the speed of the layout method (cf. Figs.~\ref{fig:resb} in Appendix for more details). Fig.~\ref{fig:gauss_1508} is a comparison of the Gauss curves obtained for the different TFL circuit depths.

\begin{figure}
    \centering
    \includegraphics[width=0.8\columnwidth]{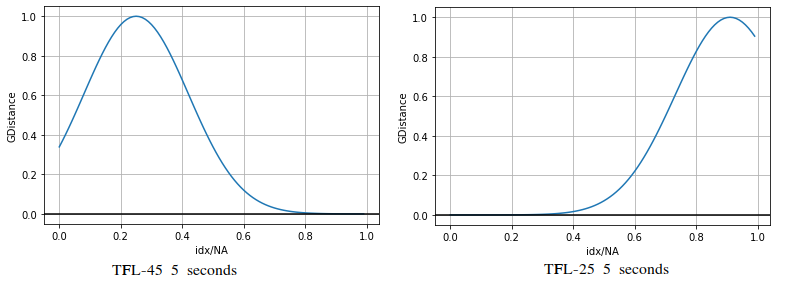}
    \caption{Two Gaussian curves obtained using WRS: left) on TFL-45 circuits with timeout 5 seconds; right) on TFL-25 circuits with timeout 5 seconds. (cf. Table~\ref{tbl:wrsbest})}.
    \label{fig:gauss_1508}
\end{figure}

\begin{figure*}
    \centering
    \includegraphics[width=0.8\textwidth]{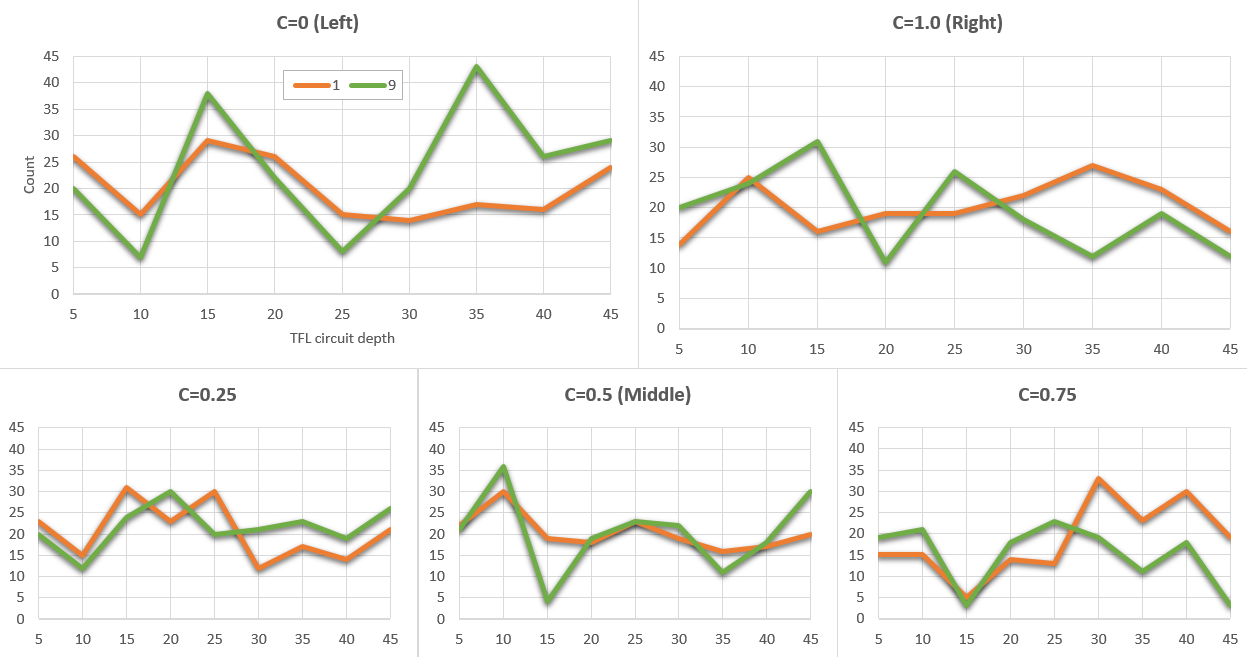}
    \caption{The $Count$ (the number of times when a parameter value was counted in the best 100 parameter combinations obtained for QXX) values to assess the influence of the Gauss center $C$ parameter on the $Ratio$. The Gauss curve is automatically adapted to circuit depth. The horizontal axis in the plots represent the different types of QUEKO TFL circuits with known depths. The vertical axis is the $Count$ value for the corresponding circuit types. There are two curves, the brown one for $MaxDepth=1$ and the green one for $MaxDepth=9$. For circuits with depths up to $30$, the green $MaxDepth=9$ performs better than brown $MaxDepth=1$ (lower $Count$ value) when the Gauss curve is positioned to the right ($C=1.0$ represents the end of the circuit) -- the last gates are more important. For circuits of depths larger than $30$, the opposite is true  -- the first gates are more important. The function $Count(P=v, |C_{TFL}|, MaxDepth)$ is the number of times when a parameter $P$ bounded to $v$ was counted in best 100 parameter combinations obtained for QXX executed for a particular value of $MaxDepth$ and for QUEKO TFL circuits $C_{TFL}$ of depth $|C_{TFL}|$. More details in the Appendix.}
    \label{fig:resc}
\end{figure*}

It is surprising that values of $C > 0$ are found, given that these assume an upfront cost of re-routing some early gates. For shallow circuits, WRS prefers $C$ values close to the end of the benchmark circuits, meaning that the last gates are more important than the others. This is in accordance with Fig.~\ref{fig:resc} where the green curve ($MaxDepth=9$) is over the orange curve ($MaxDepth=1$) for large values of $C > 0.75$ in the range of TFL depths from 5 to 30.

For deeper circuits, WRS sets the center $C$ of the Gaussian to be closer to the beginning of the circuit. This is in accordance with Fig.~\ref{fig:resc} where the vertical distance between the green and orange curves is maximum for $C < 0.25$.

Fig.~\ref{fig:resc} answers the question: Considering the different values of $MaxDepth$, where should the Gauss bell be placed relative to the start of the compiled circuit? Intuitively, this means to answer the question: Are the first gates more important than the last ones, or vice versa? Increasing values of $C$ influence the performance of QXX with decreasing $MaxDepth$ -- the orange ($MaxDepth=1$) and green ($MaxDepth=9$) curves swap positions along the vertical axis with the intersection between them being around TFl depth $30$.

\section{Conclusions}
\label{sec:conclusions}

Scalable, configurable and fast QCL methods are an imperative necessity. In the context of quantum computing clouds, continuous learning is a real possibility, because a large batch of circuits is permanently sent and executed on mainframe like machines. It is feasible to consider machine learning QCL methods for fast and accurate QCL.

We introduced QXX, a novel and parameterized QCL method. The QXX method uses a Gaussian function whose parameters determine the circuit region that influences most of the layout cost. The optimality of QXX is evaluated on the QUEKO benchmark circuits using the $Ratio$ function which expresses the factor by which the number of gates in the layed out circuit has increased. We illustrate the utility of QXX and its employed Gaussian. We show that the best results are achieved when the bell curve is non-trivially configured. QXX parameters are optimized using weighted random search (WRS). 

To increase the speed of the parameter search we train an MLP that learns QXX, and apply WRS on the resulting QXX-MLP. To crosscheck the quality of the WRS optimization and of the MLP model. This work brought empirical evidence that: 1) the performance of QXX (resulting depth $Ratio$ and speed) is on par with state of the art QCL methods; 2) it is possible to learn the QXX method parameters values and the performance degradation is an acceptable trade-off with respect to achieved speed-up compared to WRS (which per se is orders of magnitude faster than exhaustive search); 3) WRS is finding parameters values which are in accordance with the very expensive exhaustive search.

We conjecture that, in general, new cost models are necessary to improve the performance of QCL methods. Using the Gaussian function, we confirmed the observation that the cost of compiling deep circuits is determined only by some of the gates (either at the start or the end of the circuit). From this perspective the Gaussian function worked as a simplistic feature extraction. Future work will focus on more complex techniques to extract features to drive the QCL method.

\begin{acks}
AP was supported by Google Faculty Research Awards and the project NUQAT funded by Transilvania University of Bra\c sov. We are grateful to Bochen Tan for his feedback on a first version of this manuscript, explaining the QUEKO benchmarks and offering the scripts to generate and plot the presented results.
\end{acks}

\bibliographystyle{ACM-Reference-Format}
\bibliography{__main}

\appendix

\section{Background}

From an abstract point of view, register connectivity is encoded as a graph. In the simplest form possible, the graph edges are not weighted. The graph edges are unique tuples of circuit registers $(q_i, q_j)$

\begin{figure}[h!]
\centering
\includegraphics[width=4cm]{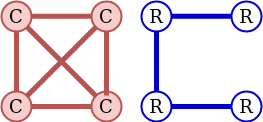}
\caption{The quantum circuit (red) has to be executed on the quantum device (blue). The circuit uses four qubits (vertices marked with C) and the hardware has four registers (blue vertices), too. The circuit assumes that operations can be performed between arbitrary pairs of registers (the edges connecting the registers). The device supports operations only between a reduced set of register pairs.}
\label{fig:map}
\end{figure}

For example, in Fig.~\ref{fig:map} the register connectivity of the circuit $C_{in}$ is the red graph. The unique tuples of device registers are the edges of the device graph. For example, in Fig.~\ref{fig:map}, the device register connectivity is the blue graph. 

Fig.~\ref{fig:gaussweight}a includes a quantum circuit example: the qubits are represented by horizontal wires, the two qubit gates are vertical lines, the control qubit is marked with a $\bullet$, and the target with $\oplus$. The red graph from Fig.~\ref{fig:map} is obtained by replacing all wires from Fig.~\ref{fig:gaussweight} with vertices and the CNOTs with edges. 

The aim of \textit{parameter optimization} is to find the parameters of a given model that return the best performance of an objective function evaluated on a validation set. In simple terms, we want to find the model parameters that yield the best score on the validation set metric.

In machine learning, we usually distinguish between the training parameters, which are adapted during the training phase, and the hyperparameters (or meta-parameters), which have to be specified before the learning phase \cite{Andonie2019}. In our case, since we do not train (adjust) inner parameters on specific training sets, we have only hyperparameters, which we will simply call here \emph{parameters}.

Parameter optimization may include a budgeting choice of how many CPU cycles are to be spent on parameter exploration, and how many CPU cycles are to be spent evaluating each parameter choice. Finding the ``best'' parameter configuration for a model is generally very time consuming. There are two inherent causes of this inefficiency: one is related to the search space, which can be a discrete domain. In its most general form, discrete optimization is NP-complete. The second cause is the evaluation of the objective function can be also expensive. We call this evaluation for one set of parameter values a \textit{trial}.

There are several recent attempts to optimize the parameters of quantum circuits. Machine learning optimizers tuned for usage on NISQ devices were recently reviewed by Lavrijsen \emph{et al} \cite{Lavrijsen2020}. Several state-of-the-art gradient-free optimizers were compared, capable of handling noisy, black-box, cost functions and stress-test them using a quantum circuit simulation environment with noise injection capabilities on individual gates. Their results indicate that specifically tuned optimizers are essential to obtaining valid results on quantum hardware. Parameter optimizers have a range of applications in quantum computing, including the Variational Quantum Eigensolver and Quantum Approximate Optimization algorithms. However, this approach has the same weaknesses like classical optimization -- global optima are exponentially difficult to achieve \cite{mcclean2018barren}. 

Currently, the most common parameter optimization approaches are \cite{Bergstra2011, Florea2019, Andonie2019, Andonie2020}: Grid Search, Random Search, derivative-free optimization (Nelder-Mead, Simulated Annealing, Evolutionary Algorithms, Particle Swarm Optimization), and Bayesian optimization (Gaussian Processes, Random Forest Regressions, Tree Parzen Estimators, etc.). Many software libraries are dedicated to parameter optimization, or have parameter optimization capabilities: BayesianOptimization, Hyperopt-sklearn, Spearmint, Optunity, etc \cite{Florea2019, Andonie2019}. Cloud based highly integrated parameter optimizers are offered by companies like Google (Google Cloud AutoML), Microsoft (Azure ML), and Amazon (SageMaker).

\section{Technical Details}

\subsection{Weighted Random Search for parameter optimization}

There are two computational complexity aspects which have to be addressed in order to find good QXX parameters: \textit{a)} Reduce the search space and implicitly the number of trials; and \textit{b)} Reduce the execution time of each trial.  In general, the performances of a parameter optimizer is determined by \cite{Andonie2019, Andonie2020}:
\begin{itemize}
	\item F1. The execution time of each trial. 
	\item F2. The total number of trials -- search space size.
	\item F3. The performance of the search. 
\end{itemize}

Search space reduction (F2) and search strategy (F3) are inter-connected and can be addressed in a sequence:  F2 is a quantitative criterion (how many). For instance, we can first reduce the number of parameters (F2), and create this way more flexibility in the following stage for F3. In this work, we do not reduce the number of parameters.

F3 is a qualitative criterion (how ``smart''). For instance, (F3), we can first rank and weight the parameters based on the functional analysis of the variance of the objective function, and then reduce the number of trials (F2) by giving more chances to the more promising trials.

There is a trade-off between F2 and F3. To address these issues and reduce the search space, we use the following standard techniques:
\begin{itemize}
	\item \textit{Instance selection:} reduce the dataset based on statistical sampling (relates to F1). 
	\item \textit{Feature selection} (relates to F1).
	\item \textit{Parameter selection}: select the most important parameters for optimization (relates to F2 \& F3).
	\item \textit{Parameter ranking}: detect which parameters are more important for the model optimization and weight them (relates to F3 \& F2).
	\item \textit{Use additional objective functions:} number of operations, optimization time, etc. (relates to F3 \& F2).
\end{itemize}

On average, the WRS method WRS converges faster than RS \cite{Florea2019}.  WRS outperformed several state-of-the art optimization methods: RS, Nelder-Mead, Particle Swarm Optimization, Sobol Sequences, Bayesian Optimization, and Tree-structured Parzen Estimator  \cite{Andonie2020}. 

In the RS approach, parameter optimization translates into the optimization of an objective function $F$ of $d$ variables by generating random values for its parameters and evaluating the function for each of these values \cite{Bergstra2011}. The function computes some quality measure or score of the model (e.g., accuracy), and the variables correspond to the parameters. The assumption is to maximize $F$ by executing a given number of trials. 

Focusing on factor F3, the idea behind the WRS algorithm is that a subset of candidate values that already produced a good result has the potential, in combination with new values for the remaining dimensions, to lead to better values of the objective function. Instead of always generating new values (like in RS), the WRS algorithm uses for a certain number of parameters the so far best obtained values. The exact number of parameters that actually change at each iteration is controlled by the probabilities of change assigned to each parameter. WRS attempts to have a good coverage of the variation of the objective function and determines the parameter importance (the weight) by computing the variation of the objective function.

\subsection{Training QXX-MLP}
\label{sec:mlp}

For the training and validation stages, we had 12 input features: circuit features (extracted using the Python package networkx) -- $max\_page\_rank$\footnote{Ranking nodes based on the structure of the incoming links.}, $nr\_conn\_comp$, $edges$, $nodes$,  $efficiency$\footnote{The efficiency of a pair of nodes in a graph is the multiplicative inverse of the shortest path distance between the nodes.},  $smetric$\footnote{The sum of the node degree products for every graph edge.} -- merged with QXX's parameters -- $MaxDepth$, $MaxChildren$, $B$, $C$, $MovementFactor$, $EdgeCost$. We have chosen PageRank, Smetric and the other metrics, in order to capture as much information about the circuits. The more features used for learning, the better the trained model is. The value to be predicted by the models was the $Ratio$ between the depth of the known optimal circuit and resulting circuit.

The performance of KNN, RF and MLP was assessed through tenfold cross validation (CV) over the whole dataset. In tenfold CV the available dataset resulted in exhaustive search is split in 10 folds. Each fold is used in turn as a validation subset, and the other nine folds are used for training. Finally, the ten performance scores obtained on the validation subsets are averaged and used as an estimation of the model's performance. 

For each of the ten train/validation splits, the optimal values of their specific hyperparameters were sought via grid search, using fivefold CV; the metric to be optimized was mean squared error. The models' specific hyperparameters and candidate values are given in Table~\ref{tbl:hps}. As both KNN and MLP are sensitive to the scales of the input data, we used a scaler to learn the ranges of input values from the train subsets; the learned ranges were subsequently used to scale the values on both train and validation subsets. We used the reference implementations from scikit--learn \cite{scikit-learn}, version 0.22.1. Excepting the hyperparameters in Table \ref{tbl:hps}, the hyperparameters of all other models are kept to their defaults.

\begin{table}[t]
\small
\centering
\begin{tabular}{l l l}
Model   & Hyperparameter            & Values \\
\hline
KNN     & Neighbors sought          & $\{2, \dots, 8\}$ \\
KNN     & Minkowski metric's $p$    & $\{1, 2\}$ \\
\hline
MLP     & Hidden layer's size       & $\{3, 10, 20, 50, 100\}$ \\
MLP     & Activation function       & ReLU, tanh \\
\hline
RF      & Maximum depth of a tree   & $\{2, 3, 4, 5, auto\}$ \\
RF      & Number of trees           & $\{2, 5, 10, 20\}$
\end{tabular}
\caption{Hyperparameter names and candidate values}
\label{tbl:hps}
\end{table}

The lowest average values for mean squared error were obtained by RF, closely followed by MLP and KNN. From RF and MLP we preferred the latter due to its higher inference speed and smaller memory footprint. 

The final MLP model was prepared by doing a final grid search for the optimal hyperparameters from Table \ref{tbl:hps}, choosing the best model through fivefold CV.  
The resulted networks look as follows: the input layer has 12 nodes, fully connected with the (only) hidden layer which hosts 100 neurons; furthermore, this layer is fully connected with the output neuron. ReLU and identity were used as activation functions for the hidden and output layers, respectively. For the hidden layer, both the number of neurons and the activation function were optimized through grid search.

\section{Evaluation}

We use the exhaustive search raw data and introduce metrics to evaluate the parameter importance of $GDepth$. The function has six parameters (see Section~\ref{sec:initial}), and we analyzed their \emph{individual} importance using WRS (see Subsection~\ref{sec:wrs}). 

For example, Table~\ref{tbl:params} lists the importances (weights) of the individual QXX parameters. These weights were computed under the strong (na\"ive) independence assumption between the parameters. Usually, parameters are statistically correlated, and we prefer a finer grained understanding of the QXX's performance. 

To compare how parameter pair influence the $Ratio$ function, we introduce two metrics called $Count$ and $Rank$. To compute these metrics we execute the exhaustive search for the three values of $MaxDepth$ and consider all the parameter configurations from Table~\ref{tbl:params} -- a six-dimensional grid search.

For a given value of $MaxDepth$ and a parameter configuration (all other five parameters) we average the resulting depth of the compiled circuit, $C_{out}$, over the circuits existing in the $TFL$ benchmark. From the total $1485$ averages, we sample the lowest 100 values, leading to an approximate 7\% sampling rate, $\frac{100}{1485}\approx 0.067$, from the total number of parameter configurations. 

The function $$Count(P=v, |C_{TFL}|, MaxDepth)$$ is the number of times when a parameter $P$ bounded to $v$ was counted in the best 100 parameter combinations obtained for QXX executed for a particular value of $MaxDepth$ for QUEKO TFL circuits $C_{TFL}$ of depth $|C_{TFL}|$. For example, $Count(B=0, 1, 30)$ is the number of parameter configurations where $B=0$ and QXX was used with a search tree of $MaxDepth=30$ to lay out TFL circuits of depth $1$. The $Count$ function can compare how, for different circuit depths, $MaxDepth$ influences the optimal values of $P$.

The $Rank$ function aggregates how different values of $P$ are ranked against each other when considering different TFL circuit depths: for the same $|C_{TFL}|$, higher rank values are better. The $Rank$ is used to suggest parameter value ranges.%
{\small%
\begin{align}
Rank(P, |C_{TFL}|) &= \sum_{i=0}^2 Count(P, |C_{TFL}|, MaxDepth_i)\\
& MaxDepth_i \in \{1, 5, 9\} \nonumber
\end{align}}

\subsection{Optimum parameters vs. circuit depth}

To speed-up QXX, we are interested in finding parameter values that keep execution times as low as possible without massively impacting the obtained $Ratio$ values. According to WRS, $MaxDepth$ is one of the most important parameters, but it is not immediately obvious if it is possible to achieve optimal $Ratio$ values using low $MaxDepth$ values. In the following, we form parameter pairs between $MaxDepth \in \{1, 5, 9\}$ and the other five QXX parameters.

The best layouts are obtained for: a) large values of $MovementFactor$, and, b) for shallow circuits, the $EdgeCost$ has to be preferably low. These observations explain the \texttt{StochasticSwap} gate scheduling method (Fig.~\ref{fig:procedure}). The $MovemenFactor$ value shows that the scheduler prefers to move a single qubit on the coupling graph. 

The way how $EdgeCost$ values are influenced by the TFL depth indicates that there exists a relation between the number of gates in the circuit and the number of edges in the coupling graph. This relation could be modelled through a density function like $\frac{nr\_gates}{nr\_edges}$. To the best of our knowledge, the effect of this density function on the coupling graph edge weights has not been investigated in the literature by now.

\begin{figure}[h!]
\centering
\includegraphics[width=8cm]{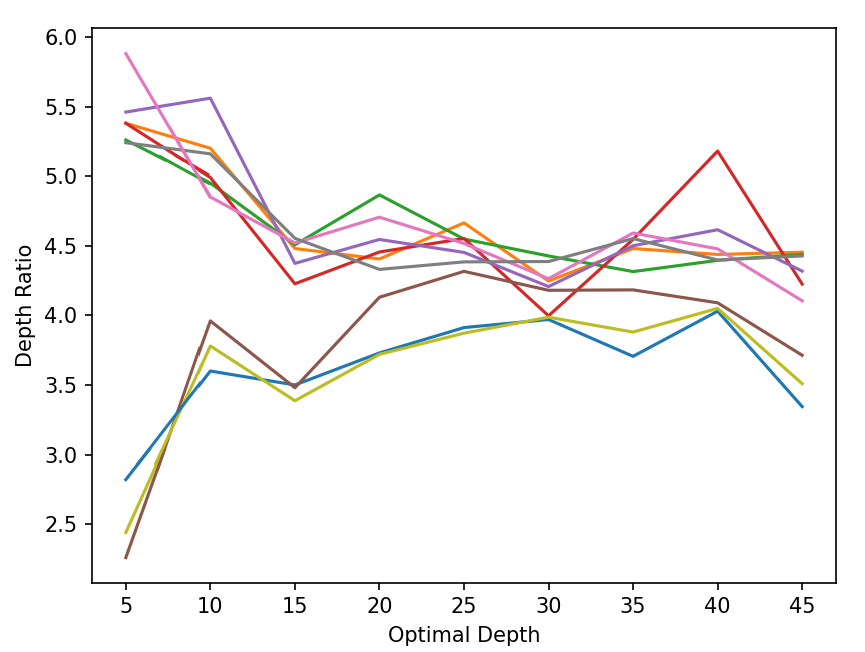}
\caption{Random parameter configurations and their influence on TFL circuit depth optimality. The axes have the same interpretation like in Fig.~\ref{fig:total}. Each line corresponds to a random parameter value configuration.
}
\label{fig:tfl}
\end{figure}

\begin{figure}[h!]
\centering
\includegraphics[width=8cm]{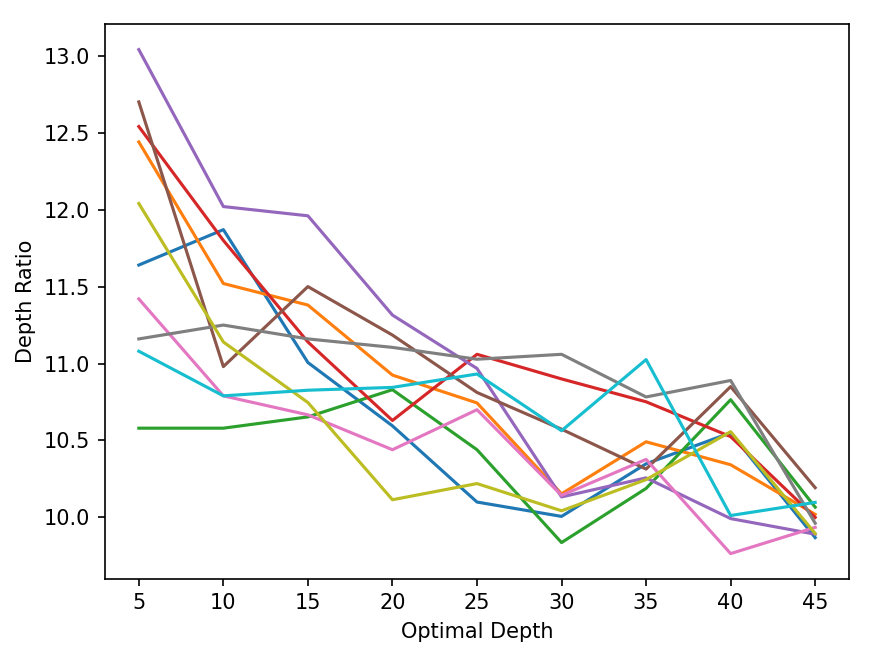}
\caption{Random parameter configurations and their influence on QSE (supremacy experiment) circuit depth optimality. The axes have the same interpretation like in Fig.~\ref{fig:total}. Each line corresponds to a random parameter value configuration.
}
\label{fig:qse}
\end{figure}

Figs.~\ref{fig:tfl} and \ref{fig:qse} show how randomly chosen parameter configurations influence the depth $Ratio$ of shallow circuits with depths up to 25.

\begin{figure*}
    \centering
    \includegraphics[width=0.8\textwidth]{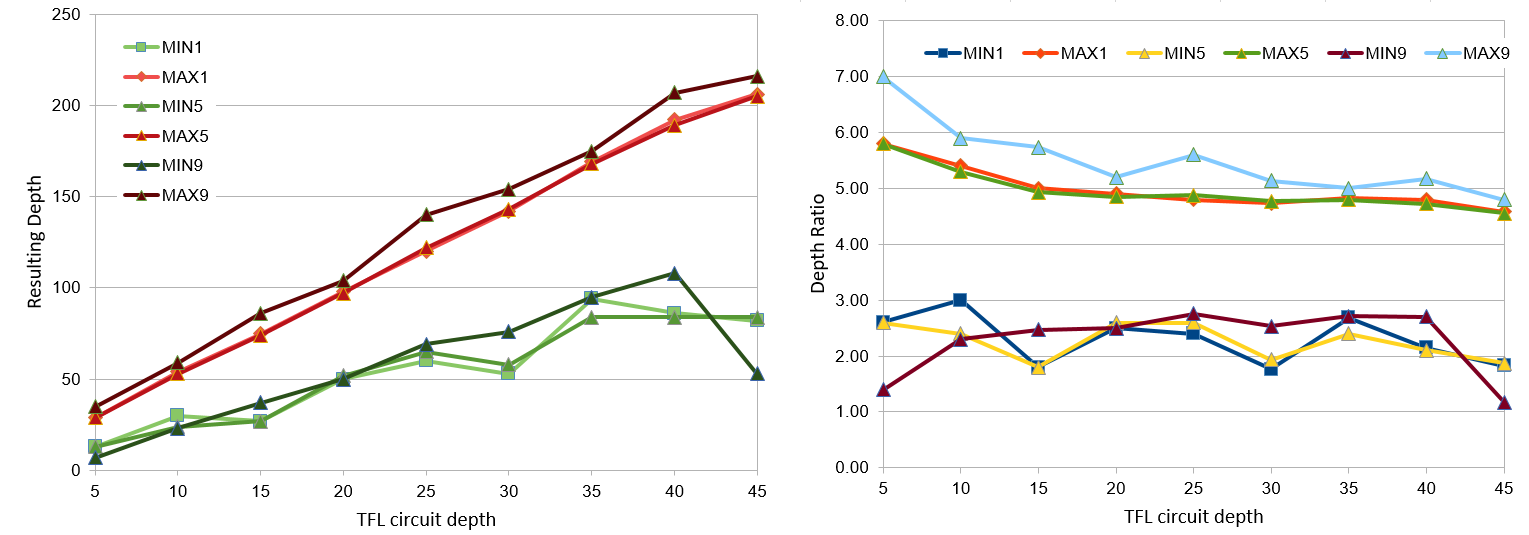}
    \caption{It should be possible to find an optimal parameter configuration irrespective of the value of $MaxDepth$: Plotting the best depth (left) and $Ratio$ (right) obtained per TFL circuit depth and $MaxDepth$ parameter. The red (MAX1, MAX5, MAX9) and green (MIN1, MIN5, MIN9) curves are the highest and lowest depths achieved for each $MaxDepth$ value (1,5,9). 
    }
    \label{fig:resdepth}
\end{figure*}

\begin{figure*}
    \centering
    \includegraphics[width=0.8\textwidth]{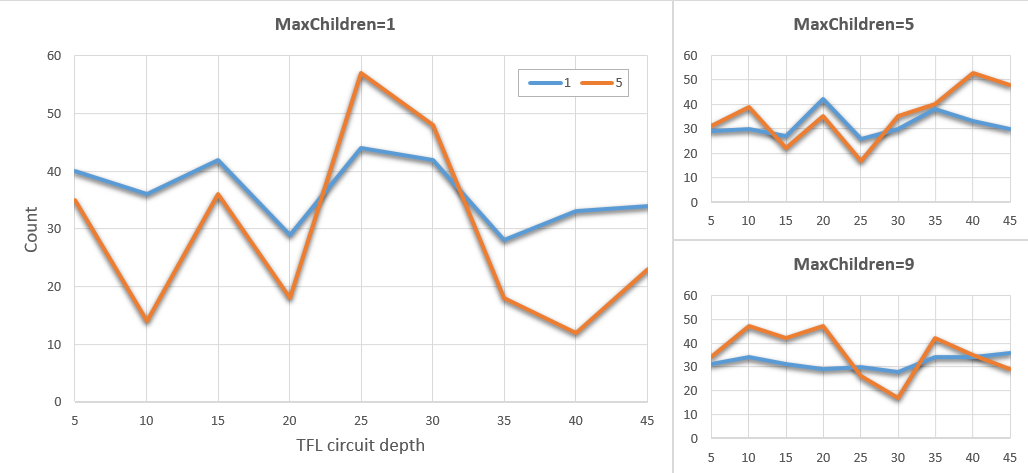}
    \caption{Exhaustive Search: The $Count$ values to assess the influence of the $MaxChildren$ parameter on the $Ratio$. For this we plot the $Count$ obtained for two values of $MaxDepth$: blue for $1$, and orange for $5$. Due to the high number of timeouts during the exhaustive search with $MaxDepth=9$, the corresponding curve was not plotted. For $MaxChildren=1$, almost for all circuits (with the exception of depth 20, 25 and 30) the best $Ratios$ is obtained for $MaxDepth=1$. As $MaxChildren$ is increased, the better results are achieved by larger $MaxDepth$ -- the orange curve is above the blue one. }
    \label{fig:reschildren}
\end{figure*}

\begin{figure*}
    \centering
    \includegraphics[width=0.8\textwidth]{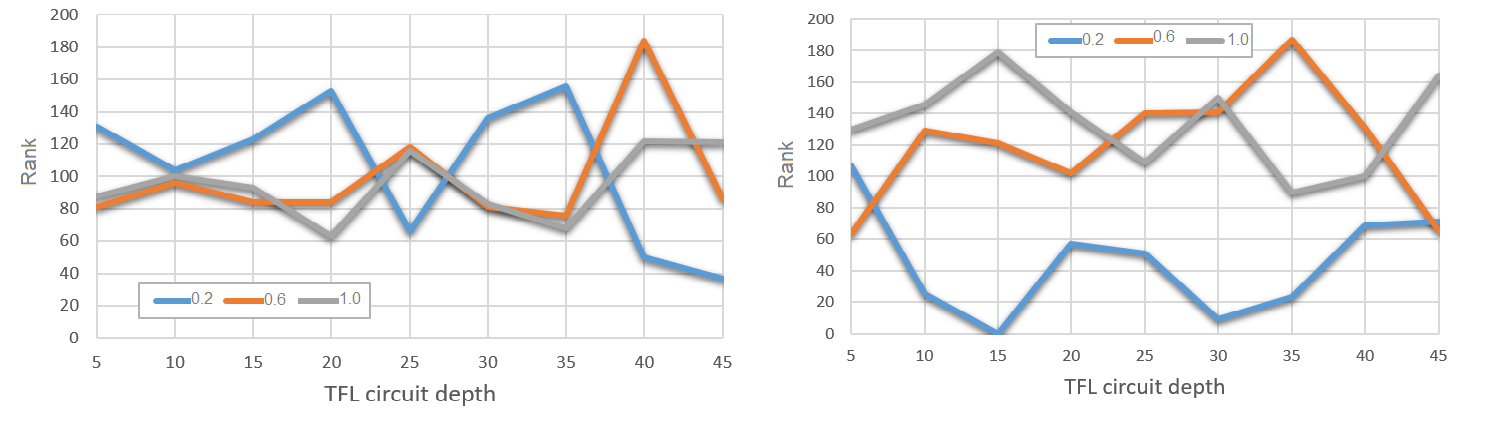}
    \caption{Exhaustive Search: left) $Rank$ values for the $EdgeCost$ parameter -- up to circuits of depths 35 QXX achieves better $Ratio$ values for edge costs of 0.2, while for deeper circuits a higher value of the edge cost delivers better $Ratio$ values; right) $Rank$ values for the $MovementFactor$ parameter -- higher parameter values perform significantly better than lower values, meaning that, cf. Fig.~\ref{fig:move}, it is better to move a single qubit instead of two across the graph.}
    \label{fig:rescx}
\end{figure*}

Fig.~\ref{fig:resdepth} (optimal parameter configuration irrespective of the value of $MaxDepth$) is supported by the results from Figs.~\ref{fig:reschildren},~\ref{fig:resb}(discussed in the next section -- the Gaussian influences $MaxDepth$) and~\ref{fig:rescx} -- the best layouts are obtained for: a) large values of $MovementFactor$, and, b) for shallow circuits, the $EdgeCost$ has to be preferably low.

We conjecture that the variability in Figs.~\ref{fig:reschildren},~\ref{fig:resb} and~\ref{fig:rescx} is mostly due to the correlations that exist between the search space size and the timeouts. For example, it can be seen that, with a small exception for medium-depth circuits, the best performing value of $MaxChildren$ is correlated with the one of $MaxDepth$. The number of timeouts we obtained for high values of $MaxDepth=9$ is an indication of this observation. As a conclusion, it does not seem to be necessary to increase the breadth of the search if the depth of the search tree is shallow.

\subsection{GDepth parameters}

\begin{figure*}
    \centering
    \includegraphics[width=0.8\textwidth]{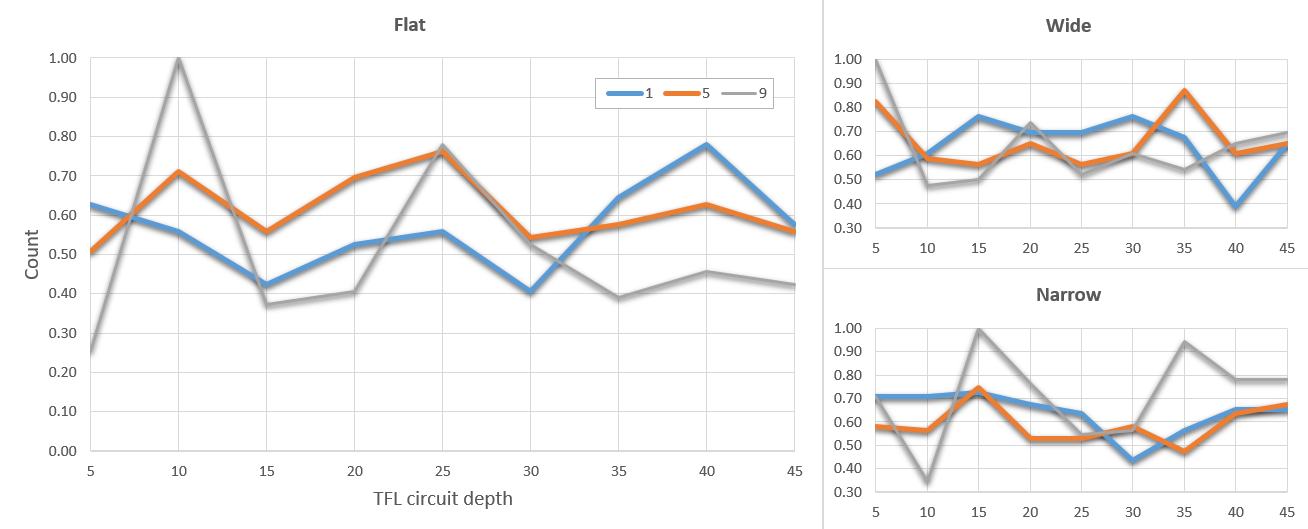}
    \caption{The value of $B$ can improve both the mapping (higher $Count$) as well as speed up the search due to lower $MaxDepth$: Normalised $Count$ values to assess the influence of the $B$ parameter on the $Ratio$. For this we plot the $Count$ obtained for three values of $MaxDepth$: blue for $1$, and orange for $5$, gray for $9$. Better results are obtained with decreasing $MaxDepth$ as the value of $B$ is increasing. For example, in the left panel, $Flat$ performs better for $MaxDepth=5$ for circuits with a depth up to $30$. Moreover, $Wide$ achieves the best depth ratios for $MaxDepth=1$. }
    \label{fig:resb}
\end{figure*}

Fig.~\ref{fig:resb} answers the question: What is the best performing value of $B$ considering the depth ($MaxDepth$) of the QXX search space? \emph{How many gates of a circuit} are important is answered by the value of $B$. Out of the 11 used values (cf. Table~\ref{tbl:params}), the first four ($0.0, 0.2, 0.4, 0.6$) are considered being $Flat$, the last four $1.4, 1.6, 1.8, 2.0$ are $Narrow$. The remaining three values are $Wide$. Due to these ranges, the values on the vertical axis are normalised to $1$. The width of the bell curve from Fig.~\ref{fig:gauss} is configurable and indicative of which circuit gates are the most important wrt. $Ratio$ optimality.

The number of preferred gates seems to be a function of the used timeout. The more time spent searching for optimal parameters, the thinner the Gaussian bell. As observed in Table~\ref{tbl:wrsbest} and Table~\ref{tbl:timeouts}, the number of timeouts for $MaxDepth=9$ is high, such that the $B$ values for timeout at 20 seconds seem not to obey the scaling observed for the other timeouts. This confirms the results of the exhaustive search as presented in Fig.~\ref{fig:resb} in Appendix, where the curve for $MaxDepth=9$ has a high variation along the vertical axis.

\begin{figure}[h!]
\centering
\includegraphics[width=8cm]{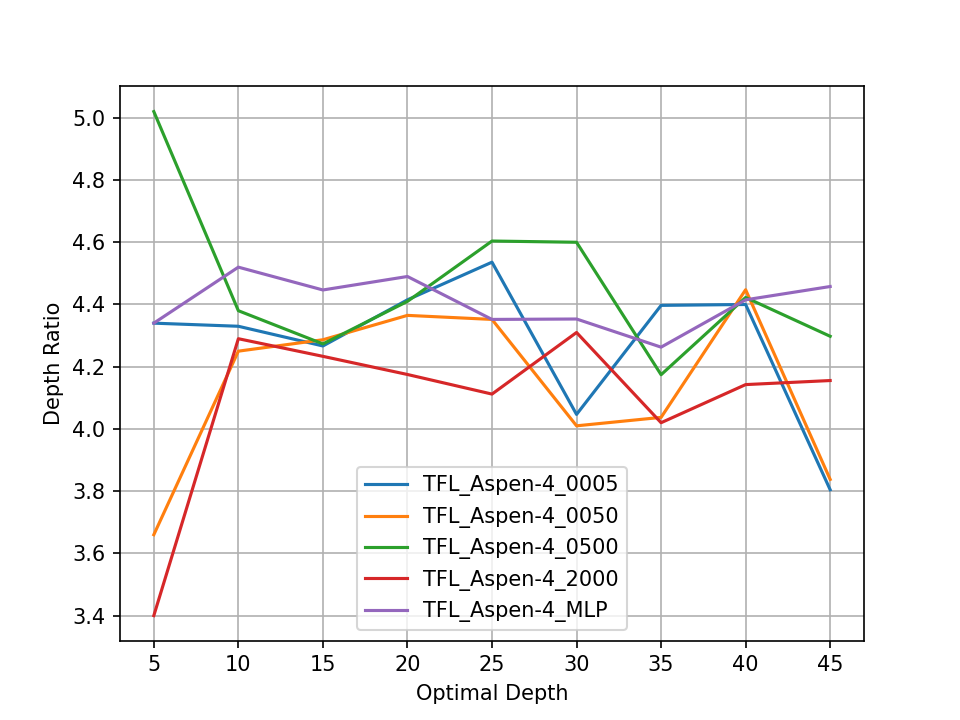}
\caption{Precision: Laying out TFL circuits with depths up to 45.}
\label{fig:timeout_tfl45}
\end{figure}

This diagram in Fig.~\ref{fig:timeout_tfl45} is similar to the one from Fig.~\ref{fig:timeout_tfl25}, but the WRS was executed on all benchmarks -- parameters were chosen to be compatible with a much larger range of circuits. The curves do not seem to converge as well as they did for the depth 25 circuits.

\end{document}